# Emergence of charge density wave domain walls above the superconducting dome in TiSe$_2$

Y. I. Joe[1], X. M. Chen[1], P. Ghaemi[1], K. D. Finkelstein[2], G. A. de la Peña[1], Y. Gan[1], J. C. T. Lee[1], S. Yuan[1], J. Geck[3], G. J. MacDougall[1], T. C. Chiang[1], S. L. Cooper[1], E. Fradkin[1], P. Abbamonte[1,4],*

[1] *Department of Physics and Frederick Seitz Materials Research Laboratory, University of Illinois, Urbana, IL 61801, USA*
[2] *Cornell High Energy Synchrotron Source, Cornell University, Ithaca, NY, 14853*
[3] *Leibniz Institute for Solid State and Materials Research, Helmholtzstrasse 20, D-01171 Dresden, Germany*
[4] *Advanced Photon Source, Argonne National Laboratory, Argonne, IL, 60439*

**Superconductivity (SC) in so-called "unconventional superconductors" is nearly always found in the vicinity of another ordered state, such as antiferromagnetism, charge density wave (CDW), or stripe order. This suggests a fundamental connection between SC and fluctuations in some other order parameter. To better understand this connection, we used high-pressure x-ray scattering to directly study the CDW order in the layered dichalcogenide TiSe$_2$, which was previously shown to exhibit SC when the CDW is suppressed by pressure [1] or intercalation of Cu atoms [2]. We succeeded in suppressing the CDW fully to zero temperature, establishing for the first time the existence of a quantum critical point (QCP) at $P_c$ = 5.1 ± 0.2 GPa, which is more than 1 GPa beyond the end of the SC region. Unexpectedly, at P = 3 GPa we observed a reentrant, weakly first order, incommensurate phase, indicating the presence of a Lifshitz tricritical point somewhere above the superconducting dome. Our study suggests that SC in TiSe$_2$ may not be connected to the QCP itself, but to the formation of CDW domain walls.**

* Email: abbamonte@mrl.illinois.edu

The term "unconventional superconductor" once referred to materials whose phenomenology does not conform to the Bardeen-Cooper-Schrieffer (BCS) paradigm for superconductivity. It is now evident that, by this definition, the vast majority of known superconductors are unconventional, notable examples being the copper-oxide, iron-arsenide, and iron-selenide high temperature superconductors, heavy Fermion materials such as $CeIn_3$ and $CeCoIn_5$, ruthenium oxides, organic superconductors such as $\varkappa$-(BEDT-TTF)2X, filled skutterudites, etc.

Despite their diversity in structure and phenomenology, the phase diagrams of these materials all exhibit the common trait that superconductivity (SC) resides near the boundary of an ordered phase with broken translational or spin rotation symmetry. For example, SC resides near antiferromagnetism in $CeIn_3$ [3], near a spin density wave in iron arsenides [4], orbital order in ruthenates [5], and stripe and nematic order in the copper-oxides [6]. The pervasiveness of this "universal phase diagram" suggests that there exists a unifying framework - more general than BCS - in which superconductivity can be understood as coexisting with some ordered phase, and potentially emerging from its fluctuations.

A classic example is the transition metal dichalcogenide family, $MX_2$, where M=Nb, Ti, Ta, and X=Se, S, which exhibit a rich competition between superconductivity and Peierls-like charge density wave (CDW) order [7]. A recent, prominent case is $1T$-$TiSe_2$, which under ambient pressure exhibits CDW order below a transition temperature $T_{CDW}$ = 202 K [8]. This CDW phase can be suppressed either with intercalation of Cu atoms [2,9], or through the application of hydrostatic pressure [1,10], causing SC to emerge. These studies suggest that the emergence of SC coincides with a quantum critical point (QCP) at which $T_{CDW}$ goes to zero, suggesting that $TiSe_2$ exemplifies the universal phenomenon of superconductivity emerging near the suppression of an ordered state. However, until now, there has been no direct observation of the QCP, impeding efforts to understand the relationship between SC and the suppression of CDW order.

A better understanding of this universal relationship requires a direct study of the CDW order parameter itself, through and beyond the QCP. Such a study would define the phenomenology and set constraints on what type of theory is needed to explain this pervasive phenomenon. For this purpose, we performed x-ray scattering experiments on the CDW order in $TiSe_2$ at low temperature and under hydrostatic pressure. Pressure is somewhat preferred to Cu doping, since the former does not introduce nonstoichiometric disorder, allowing detection of quantum critical behavior with the material nominally in its pristine state.

The origin of the CDW order in $TiSe_2$ has been the subject of debate. Explanations for its existence range from traditional Fermi surface nesting [7] to an excitonic insulator scenario [11] to an indirect Jahn-Teller effect [12], with the relative importance of electron-electron and electron-phonon interactions still in dispute [13,14,15,16,17]. What is clear is that the CDW in $TiSe_2$ is a condensate of electron-hole pairs with nonzero total momentum, in which electron-electron interactions, the Peierls distortion, and lattice pinning effects all play a prominent role.

X-ray experiments were carried out at the Cornell High Energy Synchrotron Source (CHESS) on $TiSe_2$ crystals mounted in a screw-driven diamond anvil cell (DAC) in a closed-cycle cryostat. Experiments were done without an analyzer, the detector integrating over all scattered photon energies. In this case, the experiment measures an energy-integrated, equal time correlation function, $S(\mathbf{q})$, which is the Fourier transform of the time-averaged electron density correlator, incorporating both classical and quantum fluctuations [18]. The lowest accessible temperature was 8.3 K, which was not low enough to enter the SC phase but was adequate to address the energy scale of the CDW, even close to the QCP.

The signature of a CDW in TiSe$_2$ is the existence of an order parameter reflection at wave vector of (1/2, 1/2, 1/2), where the Miller indices ($H,K,L$) represent a momentum transfer **q** = $H$ **b**$_1$ + $K$ **b**$_2$ + $L$ **b**$_3$, where **b**$_1$, **b**$_2$, **b**$_3$ are basis vectors of the reciprocal lattice. We note that, unlike most other dichalcogenides such as NbSe$_2$ or TaS$_2$, the CDW in TiSe$_2$ is commensurate at ambient pressure, with $H$, $K$, and $L$ always being rational fractions [8,19,16], suggesting that lattice pinning contributes significantly to its formation energy.

In confirmation of the conclusions of ref. [1,10], the CDW was found to be suppressed by hydrostatic pressure (Fig. 1a). The $T_{CDW}$ value was found to fall to zero, indicating the presence of a quantum critical point at a pressure of $P_c$ = 5.1 ± 0.2 GPa. The pressure dependence of $T_{CDW}$ can be fit well using a single power law, i.e., $T_{CDW}(P) = |P-P_c/P_c|^\beta$, where β= 0.87±0.08 is an effective exponent characterizing the observed order parameter suppression (Fig. 1c). If one assumes that a Hertz-Millis picture of quantum critical scaling applies near the QCP [20,21, this exponent β = $v\,z$, where $v$ is the exponent of the correlation length and $z$ is the dynamical critical exponent. In mean field theory $v$ = 1/2 and $z$ = 2 for a CDW transition, implying β = 1, which is close to the measured value.

We note that, in contrast to expectations based on refs. [2,9,11], this QCP does not reside within the superconducting dome, which spans the range 2 GPa < P < 4 GPa. This might seem to cast doubt on the presumed connection between CDW fluctuations and SC. However, we will show below that the situation is more subtle.

Surprisingly, a trace amount of CDW correlations were observed beyond the QCP. At P = 5.9 GPa a weak, resolution-limited CDW reflection is still present, whose integrated intensity is 10$^4$x smaller than at ambient pressure. This temperature-independent feature was found to be present everywhere in the normal state of the *PT* phase diagram, even above the classical transition at *P* = 0 (Fig. 1b). This suggests that our system is experiencing a small, uniform, symmetry-breaking field, perhaps due to nonuniform stresses in the pressure cell (see Supporting Material). We will discuss this effect further below.

Just as $T_{CDW}$ is suppressed by pressure, so too is the low-temperature value of the CDW integrated intensity, $I(T=0)$. This provides insight into the pressure dependence of the interactions that drive the CDW. According to the Landau theory of McMillan [22], the ratio $I(T=0)/T_{CDW}$ is proportional to the coupling constant, $g$, which we display in Fig. 1c. Surprisingly, $g$ rises continuously with pressure up to the QCP, i.e., the magnitude of the CDW coupling increases as the order itself is suppressed. This suggests that the suppression by pressure occurs by some mechanism other than a reduction in the strength of the interactions. The complete phase diagram, showing the order parameter intensity, phase boundary, and the location of the QCP, is shown in Fig. 2.

Near the QCP one might expect to observe quantum critical fluctuations, which should appear near the transition as power law tails in the correlation function, i.e., $S(q) \propto q^{-(2-\eta)}$, where $q$ is measured with respect to the ordering wave vector. However, unlike previous studies of the classical, second-order transition, in which such fluctuations are clearly visible [19], no tails were observed in our study of the quantum transition, even at high momentum resolution and very close to the phase boundary. This is most likely a consequence of the small, internal strain gradient in the pressure cell, which for the case of a CDW can act as a uniform, symmetry breaking field, suppressing fluctuations near the transition.

Our study sheds light on the mechanism by which pressure suppresses the CDW in TiSe$_2$. Because the coupling constant increases with pressure, a different contribution to the free energy must be responsible for the suppression of CDW order, such as the lattice stiffness [23] or the degree of commensuration, which contributes through the pinning term in the free energy [22]. In particular, because the CDW in TiSe$_2$ has always been observed to be commensurate, evidence for incommensurate behavior would be a major clue as to how the order is suppressed and, by association, why superconductivity forms.

To evaluate this possibility, we employed a novel, precision technique for measuring the degree of commensurability of the CDW. The difficulty in such a measurement is referencing the wave vector of the CDW to that of a Bragg reflection of the underlying lattice. In the case of TiSe$_2$, there is no Bragg reflection close to the CDW, so any small geometric misalignment propagates into large errors in the values of ($H$, $K$, $L$).

To overcome this difficulty, we performed scattering experiments at two photon energies simultaneously. An x-ray source with diffractive optics always contains harmonics, i.e., a small percentage of photons with precisely half the wavelength, in our case a tiny fraction of 36 keV photons in our nominally 18 keV beam. Rather than filtering out the harmonics, we employed an energy-proportional detector to measure both energies in parallel, allowing simultaneous measurement of the ($H$, $K$, $L$) and ($2H$, $2K$, $2L$) points in momentum space. This allowed us reference *in situ* the CDW correlations near (1/2, 1/2, 1/2) to the Bragg reflection at (1,1,1), eliminating errors due to misalignment, temperature drifts, etc. This measurement was performed at a selection of points across the phase diagram, at pressures near and below the SC dome. The results are summarized in Fig. 3, with labels "C" or "IC" added to the corresponding points in Fig. 2.

Consistent with past x-ray studies [8], at low pressures the CDW was found to be commensurate to within the experimental resolution of a few thousandths of a reciprocal lattice unit (r.l.u.). This was true even of the anomalous normal state peak observed above $T_{CDW}$ (Fig. 1b). Surprisingly, however, at P = 3 GPa - directly above the superconducting dome - weak incommensurate behavior was observed at the level of $2 \times 10^{-3}$ r.l.u. At the lowest temperatures the CDW is commensurate, but IC behavior emerged as the temperature was increased, becoming most pronounced near the phase boundary at $T$ = 90 K. The incommensurability was largest along the $L$ direction (Fig. 3), suggesting that the IC behavior may be attributed to phase slips in the stacking of the CDW order along the c axis, separated by a distance of 1/0.002 = 500 lattice parameters. While it was not possible to perform this measurement at every point in the phase diagram, observation of IC behavior is sufficient to establish the existence of a Lifshitz multicritical point, analogous to behavior in helical ferromagnets [24], somewhere in the region above the SC dome.

The transition from C to IC CDW order may be weakly first order. While the effect lies near the limit of our resolution, the transition appears to exhibit a coexistence region for 25 K < T < 50 K, in which both C and IC phases are present (Fig. 3). A first order character would make this C/IC transition similar to those in other dichalcogenides, including 2$H$-NbSe$_2$ [25] and 1$T$-TaS$_2$ [26].

Our study has significant bearing on the existing picture of CDW melting and the emergence of SC in TiSe$_2$. First, the existence of an IC phase suggests that the natural wave vector of the CDW, which is set by the Fermi surface topology, shifts from the commensurate point as the energy bands adjust with pressure. This destabilizes the CDW by reducing the energy saved from the lattice pinning potential. We note that this view is not dependent on a Fermi surface nesting picture of the CDW, since the wave vector of strong coupling phases, such as the excitonic insulator, are also determined by the band topology [7].

Next, the separation of SC and the QCP by more than 1 GPa suggests that superconductivity in TiSe$_2$ is not directly associated with the vanishing of the CDW itself. Instead, we find that the SC region coincides with a commensurate / incommensurate transition, at which the CDW order acquires a dilute concentration of phase slips. This suggests that the onset of SC in TiSe$_2$ is associated with the formation of domain walls in the CDW order. This suggests, more broadly, that the connection between SC and quantum criticality may be less direct than previously supposed, involving not the fluctuations in the amplitude itself, but the quantum dynamics of domain walls [27,28].

**Methods**

*Sample preparation.* 1*T*-TiSe$_2$ single crystals were grown via an iodine vapor transport technique. Ti and slight excesses of Se powders were loaded in a vacuum sealed quartz tube along with a small amount of iodine. The tube was heated to 570-640 °C for 6 hours, maintained in a temperature gradient for a week, and cooled to room temperature over 12 hours.

*Hydrostatic pressure*. A Bell-Mao type, screw-driven diamond anvil cell was used to generate hydrostatic pressure. A 4:1 volumetric methanol-ethanol mixture was used as a pressure-transmitting medium. This mixture was chosen because it retains hydrostaticity up to 10 GPa. A 400μm thick copper gasket was pre-indented down to 230μm and drilled for the sample chamber, which was about 450μm in diameter. The pressure was calibrated by measuring the fluorescence of a small ruby chip placed in the sample beside the TiSe$_2$ crystal.

*X-Ray Experiments*. Experiments were carried out at the C1 beam line at CHESS with a 4K displex installed on a four-circle diffractometer. To simultaneously measure scattering at (0.5,0.5,0.5) and (1,1,1), an energy-proportional scintillation counter was fed into two independent single-channel analyzers to bin the 18 keV and 36 keV photons into separate scaler channels.

**Acknowledgements.**

We gratefully acknowledge helpful input from P. B. Littlewood, M. R. Norman and R. Osborn. This work was supported by the U.S. Department of Energy under Grant No. DE-FG02-07ER46453. Use of the CHESS was supported by the National Science Foundation and the National Institutes of Health/National Institute of General Medical Sciences under NSF award DMR-0936384. T. C. C. was supported by DOE BES Grant No. DE-FG02-07ER46383.

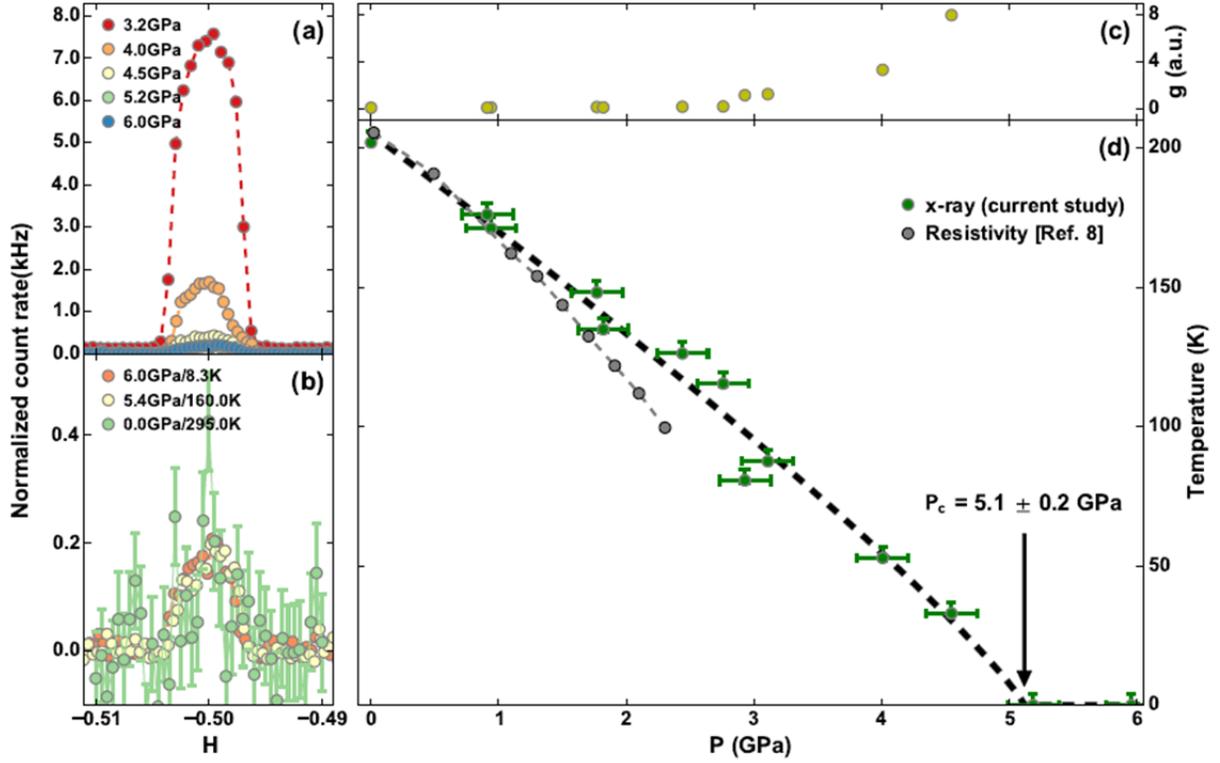

**Figure 1** Suppression of the CDW in 1T-TiSe2 with temperature and pressure. (a) Scans through the CDW ordering vector showing suppression of CDW correlations with increasing pressure. (b) Small, residual CDW correlations observed in the normal state, both at high and low pressure. (c) Pressure-dependence of the CDW coupling constant, $g$, which increases with pressure up to the phase boundary. (d) Phase boundary delineating the ordered and disordered phases in the pressure-temperature plane, showing the location of the quantum critical point at $P_c$ = 5.1 ± 0.2 GPa. This curve was found to fit well to a single power law over the entire region, $T_{CDW}(P) = |P-P_c/P_c|^\beta$, where β= 0.87±0.08, which is close to the expected mean field value (see text).

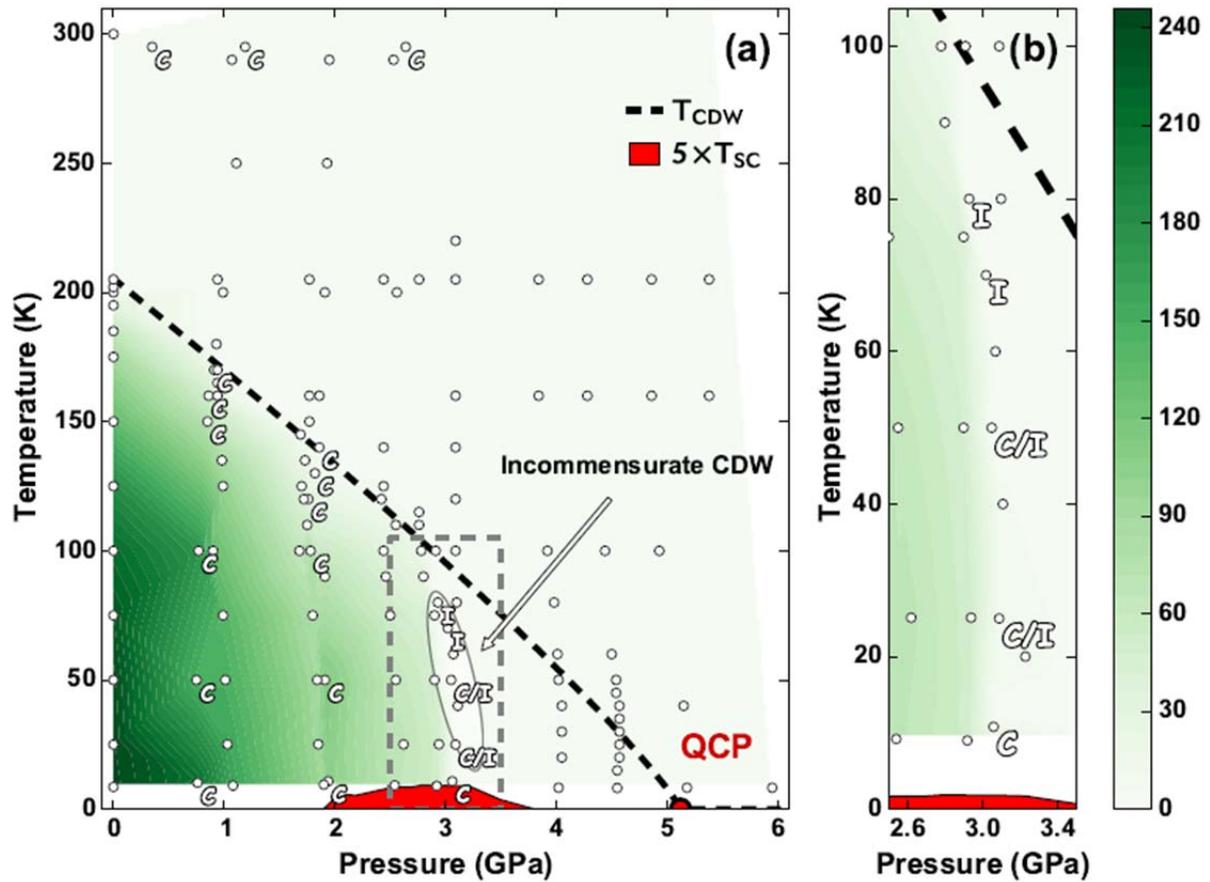

**Figure 2** Summary pressure-temperature phase diagram of TiSe$_2$. (a) Broad phase diagram showing CDW ordered, normal state, and superconducting phase boundaries. The green color scale indicates the integrated intensity of CDW correlations, including both the C and IC components. The superconducting $T_{SC}$ value has been exaggerated by a factor of 5 for visibility. Points where the precise commensurability was measured are labeled C, I, or C/I, indicating commensurate, incommensurate, or coexistence, respectively. (b) Zoom-in on the region exhibiting the transition between commensurate and incommensurate order.

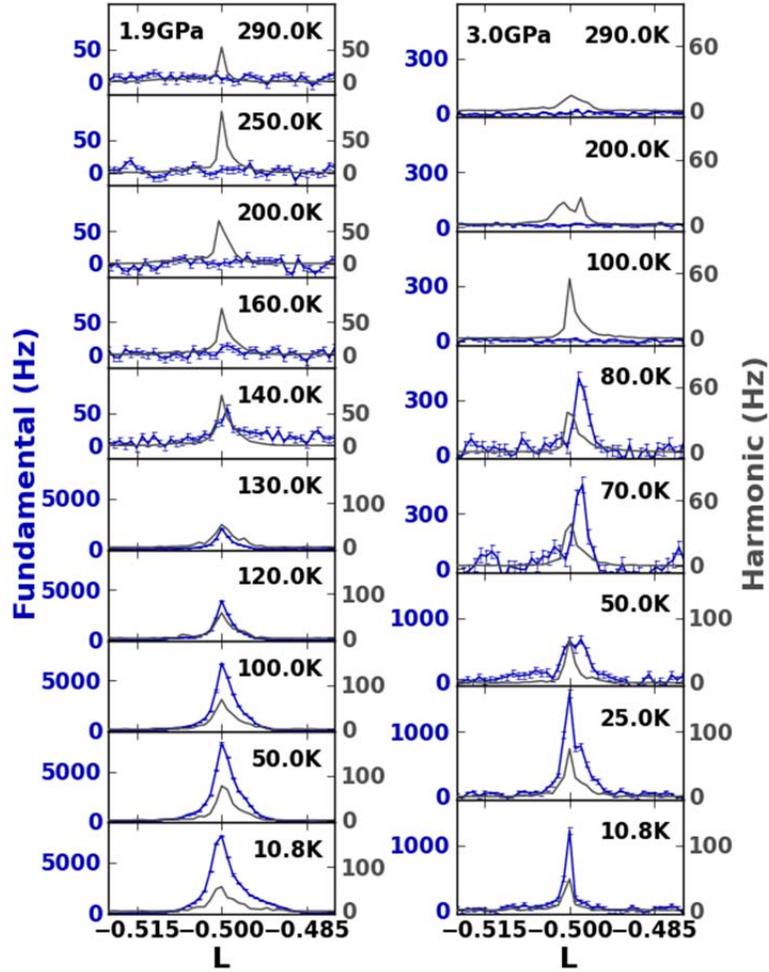

**Figure 3** Precision measurement of the CDW commensurability in TiSe$_2$ at two selected pressures, using the two-energy technique introduced in the text. Blue curves correspond to the fundamental, CDW signal near (0.5, 0.5, 0.5), and gray curves show scattering from the harmonic due to the (1, 1, 1) Bragg reflection at double the momentum. At P=1.9 GPa (left) the fundamental and harmonic curves always coincide, even up to the phase boundary at T = 160 K. However at P= 3 GPa, above the superconducting dome, the curves separate above 25 K, indicating a transition to incommensurate behavior. All scans are plotted against the z-component of the momentum, L, which is the direction of largest incommensurability.

**References**

1. A. F. Kusmartseva, B. Sipos, H. Berger, L. Forró, E. Tutiš, Pressure Induced Superconductivity in 1T-TiSe$_2$, Phys. Rev. Lett. **103**, 236401 (2009)
2. E. Morosan, H. W. Zandbergen, B. S. Dennis, J. W. G. Bos, Y. Onose, T. Klimczuk, A. P. Ramirez, N. P. Ong, R. J. Cava, Superconductivity in Cu$_x$TiSe$_2$, Nature Phys. **2**, 544-550 (2006)
3. N. D. Mathur, F. M. Grosche, S. R. Julian, I. R. Walker, D. M. Freye, R. K. W. Haselwimmer, G. G. Lonzarich, Magnetically mediated superconductivity in heavy fermion compounds, Nature **394**, 39-43 (1998)
4. P. C. Canfield, S. L. Bud'ko, FeAs-Based Superconductivity: A Case Study of the Effects of Transition Metal Doping on BaFe$_2$As$_2$, Ann. Rev. Cond. Matt. Phys. **1**, 27-50 (2010)
5. S. Nakatsuji, Y. Maeno, Quasi-Two-Dimensional Mott Transition System Ca$_{2-x}$Sr$_x$RuO$_4$, Phys. Rev. Lett. **84**, 2666 (2000)
6. J. M. Tranquada, B. J. Sternlieb, J. D. Axe, Y. Nakamura, S. Uchida, Evidence for stripe correlations of spins and holes in copper-oxide superconductors, Nature **375**, 561-563 (1995)
7. K. Rossnagel, On the origin of charge density waves in select layered transition-metal dichalcogenides, J. Phys.: Condens. Matter **23**, 213001 (2011)
8. F. J. DiSalvo, D. E. Moncton, J. V. Waszczak, Electronic properties and superlattice formation in the semimetal TiSe$_2$, Phys. Rev. B **14**, 4321-4328 (1976)
9. H. Barath, M. Kim, J.F. Karpus, S. L. Cooper, P. Abbamonte, E. Fradkin, E. Morosan, R.J. Cava, Quantum and classical mode softening near the charge-density-wave/superconductor transition of Cu$_x$TiSe$_2$, Phys. Rev. Lett. **100**, 106402 (2008)
10. C. S. Snow, J. F. Karpus, S. L. Cooper, T. E. Kidd, T.-C. Chiang, Quantum melting of the charge-density-wave state in 1T-TiSe$_2$, Phys. Rev. Lett. **91**, 136402 (2003)
11. M. M. May, C. Brabetz, C. Janowitz, R. Manzke, Charge-Density-Wave Phase of 1T-TiSe$_2$: The Influence of Conduction Band Population, Phys. Rev. Lett. **107**, 176405 (2011)
12. T. E. Kidd, T. Miller, M. Y. Chou, T.-C. Chiang, Electron-hole Coupling and the Charge Density Wave Transition in TiSe$_2$, Phys. Rev. Lett. **88**, 226402 (2002)
13. J. van Wezel, P. Nahai-Williamson, S. S. Saxena, Exciton-phonon-driven charge density wave in TiSe2, Phys. Rev B **81**, 165109 (2010)
14. C. Monney, C. Battaglia, H. Cercellier, P. Aebi, H. Beck, Exciton Condensation Driving the Periodic Lattice Distortion of 1T-TiSe$_2$, Phys. Rev. Lett. **106**, 106404 (2011)
15. S. Hellmann, T. Rohwer, M. Kalläne, K. Hanff, C. Sohrt, A. Stange, A. Carr, M.M. Murnane, H.C. Kapteyn, L. Kipp, M. Bauer, K. Rossnagel, Time-domain classification of charge-density-wave insulators, Nature Comm. **3**, 1-8 (2012)
16. J.-P. Castellan, S. Rosenkranz, R. Osborn, Q. Li, K.E. Gray, X. Luo, U. Welp, G. Karapetrov, J. P. C. Ruff, J. van Wezel, Chiral Phase Transition in Charge-Ordered 1T-TiSe$_2$, Phys. Rev. Lett. **110**, 196404 (2013)
17. F. Weber, S. Rosenkranz, J.-P. Castellan, R. Osborn, G. Karapetrov, R. Hott, R. Heid, K.-P. Bohnen, A. Alatas, Electron-Phonon Coupling and the Soft Phonon Mode in TiSe$_2$, Phys. Rev. Lett. **107**, 266401 (2011)
18. P. Abbamonte, G. C. L. Wong, D. G. Cahill, J. P. Reed, R. H. Coridan, N. W. Schmidt, G. H. Lai, Y. I. Joe, D. Casa, Ultrafast imaging and the phase problem for inelastic x-ray scattering, Adv. Mater. **22**, 1141-1147 (2010)
19. M. Holt, P. Zschack, H. Hong, M. Y. Chou, T.-C. Chiang, X-Ray Studies of Phonon Softening in TiSe$_2$, Phys. Rev. Lett. **86**, 3799 (2009)
20. J. A. Hertz, Quantum critical phenomena, Phys. Rev. B **14**, 1165-1184 (1976)
21. A. J. Millis, Effect of nonzero temperature on quantum critical points in itinerant fermion systems, Phys. Rev. B **48**, 7183-7196 (1993) 22  W. L. McMillan, Microscopic model of charge-density waves in 2H-TaSe$_2$, Phys. Rev. B **16**, 643-650 (1977)
23  M. Calandra, F. Mauri, Charge-Density Wave and Superconducting Dome in TiSe2 from Electron-Phonon interaction, Phys. Rev. Lett. **106**, 196406 (2011)
24  R. M. Hornreich, The Lifshitz point: phase diagrams and critical behavior, J. Magn. Mag. Mater. **15-18**, 387-392 (1980)
25  D.E. Moncton, J. D. Axe, F. J. DiSalvo, Study of Superlattice Formation in 2H-NbSe$_2$ and 2H-TaSe$_2$ by Neutron Scattering, Phys. Rev. Lett.  **34**, 734 (1975)
26  B. Sipos, A. F. Kusmartseva, A. Akrap, H. Berger, L. Forró, E. Tutiš, From Mott state to superconductivity in 1T-TaS$_2$, Nature Mater. **7**, 960-965 (2008)
27  E. Fradkin, S. A. Kivelson, High-temperature superconductivity: ineluctable complexity, Nature Phys. **8**, 864-866 (2012)
28  A. Mesaros, K. Fujita, H. Eisaki, S. Uchida, J. C. Davis, S. Sachdev, J. Zaanen, M. J. Lawler, Eun-Ah Kim, Topological Defects Coupling Smectic Modulations to Intra–Unit-Cell Nematicity in Cuprates, Science **333**, 426-430 (2011)